\documentclass[10pt,aps,prl,showpacs,superscriptaddress,nofootinbib,twocolumn,preprintnumbers,floatfix,showkeys]{revtex4-1}
\usepackage{graphicx,amsmath,amssymb,soul,enumitem,bbm,dcolumn}
\usepackage{booktabs,multirow,array,slashed,dsfont,kantlipsum}
\usepackage[dvipsnames]{xcolor}
\usepackage{nicefrac}
\usepackage[utf8]{inputenc}
\usepackage{wasysym}
\allowdisplaybreaks

\usepackage{hyperref}
\hypersetup{colorlinks=true,urlcolor=PineGreen,linkcolor=PineGreen,citecolor=PineGreen}

\begin{document}

\title{Three-body unitarity versus finite-volume $\pi^+\pi^+\pi^+$ spectrum from lattice QCD}

\author{M.\ Mai}
\email{maximmai@gwu.edu}
\affiliation{Institute for Nuclear Studies and Department of Physics, The George Washington University, Washington, DC 20052, USA}
\author{M.\ D\"oring}
\email{doring@gwu.edu}
\affiliation{Institute for Nuclear Studies and Department of Physics, The George Washington University, Washington, DC 20052, USA}
\affiliation{Thomas Jefferson National Accelerator Facility, Newport News, VA 23606, USA}

\author{C.\ Culver}
\email{chrisculver@email.gwu.edu}
\affiliation{Institute for Nuclear Studies and Department of Physics, The George Washington University, Washington, DC 20052, USA}

\author{A.\ Alexandru}
\email{aalexan@gwu.edu}
\affiliation{Institute for Nuclear Studies and Department of Physics, The George Washington University, Washington, DC 20052, USA}
\affiliation{Department of Physics, University of Maryland, College Park, MD 20742, USA}


\begin{abstract}
Strong three-body interactions above threshold govern the dynamics of many exotics and conventional excited mesons and baryons. Three-body finite-volume energies calculated from lattice QCD promise an ab-initio understanding of these systems. We calculate the three-$\pi^+$ spectrum unraveling the three-body dynamics that is tightly intertwined with the $S$-matrix principle of three-body unitarity and compare it with recent lattice QCD results. For this purpose, we develop a formalism for three-body systems in moving frames and apply it numerically.
\end{abstract}



\maketitle

\section{Introduction} 

The dynamics of three-body systems above threshold play a key role to our understanding of strong forces.
Many emblematic resonances exhibit  significant three-body decay channels, such as the Roper resonance $N(1440)1/2^+$ which, despite its low mass, couples strongly to the $\pi\pi N$ channel leading to a very non-standard line shape and complicated analytic structure~\cite{Arndt:2006bf, Doring:2009yv}. The $\pi\pi N$ channels play also a significant role for other excited baryons and their description needs a quantitative understanding of three-body dynamics. Similarly, axial mesons  like the $a_1(1260)$ and, supposedly, exotics decay into three particles~\cite{Alekseev:2009aa}. 

The quantitative understanding of three-body systems in terms of QCD represents a long-term goal in hadronic physics. In lattice QCD (LQCD), the Hamiltonian is discretized and its eigenvalues are determined. These numerically demanding calculations are necessarily performed in a finite volume with periodic boundary conditions.  This leads to a discrete eigenvalue spectrum in contrast to the continuous spectral density of scattering states in the infinite volume. 
These finite-volume effects are determined by hadron interactions and they offer a key to understanding these interactions {arising} from quark-gluon dynamics.

In this study, we compare the results of a recently developed infinite-volume mapping technique~\cite{Mai:2017bge} with new finite-volume energy eigenvalues~\cite{Horz:2019rrn}. 
These data are calculated with multi-pion operators allowing for the reliable extraction of energy eigenvalues, above threshold and in different irreducible representation, providing, for the first time, access to three-body dynamics from first principles. 
Similarly to the case of the $2\pi^+$ system, that represents the first physical application of the original L\"uscher formalism~\cite{Sharpe:1992pp, Kuramashi:1993ka, Gupta:1993rn, Yamazaki:2004qb, Aoki:2005uf}, the $3\pi^+$ system permits few partial waves and is an ideal system to study the pertinent finite-volume effects. This is a first step towards more complicated resonant systems that usually exhibit a complex pattern into two and three-body final states.

Recent progress in the three-particle sector is summarized in Ref.~\cite{Hansen:2019nir}, see also Ref.~\cite{Briceno:2017max} for a broader overview. In elastic two-particle scattering, each energy eigenvalue can be mapped to a phase shift~\cite{Luscher:1986pf, Luscher:1990ux}. However, the $3\to 3$ reaction has eight independent kinematic variables (not including spin.)  This requires a {new} formalism to map the discrete energy spectrum to infinite volume {quantities}. 

Scattering amplitudes cannot be directly computed as infinite-volume limits of finite-volume observables. However, even without fully resolving the three-body dynamics explicitly, methods exist that take into account the contribution of three-body states~\cite{Agadjanov:2016mao, Bulava:2019kbi, Hansen:2017mnd, Hashimoto:2017wqo}. These methods connect finite-volume data with infinite volume properties using either the optical potential or by extracting the spectral density from a correlator.

Methods resolving explicitly the three-body structure of the amplitude are being developed by different groups, for bound states~\cite{Meng:2017jgx, Hammer:2017uqm, Meissner:2014dea, Bour:2011ef, Kreuzer:2010ti} and energy levels above threshold~\cite{Guo:2019hih, Romero-Lopez:2019qrt, Mai:2018djl, Zhu:2019dho, Guo:2018ibd,  Doring:2018xxx, Guo:2018xbv, Romero-Lopez:2018rcb, Klos:2018sen,  Mai:2017bge, Guo:2017crd,  Guo:2017ism,  Hammer:2017kms, Hammer:2017uqm, Briceno:2018aml, Briceno:2018mlh, Briceno:2017tce, Guo:2016fgl, Hansen:2016fzj, Hansen:2015zga, Jansen:2015lha, Hansen:2014eka, Polejaeva:2012ut, Roca:2012rx, Briceno:2012rv, Bour:2012hn,  Kreuzer:2012sr, Kreuzer:2009jp, Kreuzer:2008bi}. The equivalence of different formalisms was  discussed recently~\cite{Jackura:2019bmu, Briceno:2019muc} (see also Refs.~\cite{Mikhasenko:2019vhk, Jackura:2018xnx}).
The $1/L$ expansion for threshold states was developed in Ref.~\cite{Detmold:2008gh} and for low-lying excited states in Ref.~\cite{Pang:2019dfe}, see also Ref.~\cite{Briceno:2018mlh}. 
A formalism for coupled two and three-body systems was developed in Ref.~\cite{Briceno:2017tce}; higher-spin two-particle sub-systems were considered in Ref.~\cite{Blanton:2019igq}.
First numerical studies~\cite{Mai:2017bge, Hammer:2017kms, Briceno:2018mlh} demonstrated the feasibility of different formalisms. 

The first application of a three-body formalism to an actual physical system  above threshold was achieved in Ref.~\cite{Mai:2018djl}. Eigenvalues for the  $3\pi^+$ system were analyzed as calculated by the NPLQCD collaboration~\cite{Beane:2007es, Detmold:2008fn}. 

One of the problems in the lattice QCD calculation of energies for channels where three-body states are relevant is the need for many-hadron type operators to reliably determine the spectrum, as demonstrated, e.g., in Ref.~\cite{Lang:2016hnn}.
Indeed, meson-baryon operators are often included in the operator
basis~\cite{ Andersen:2017una, Kiratidis:2016hda, Lang:2016hnn, Lang:2012db}.
Also, results on the Roper resonance at almost physical masses~\cite{Lang:2016hnn} suggest the need to map out finite-volume effects in two and three-body coupled channels, namely the $\pi N,\,f_0(500)N,\,\pi\Delta,\,\rho N, \ldots$ channels.

In the meson sector, lattice QCD results are available for channels where three-body states should be relevant~\cite{Lang:2014tia, Woss:2018irj, Woss:2019hse}, albeit only for pion masses and/or volumes at which the $\rho$ meson can approximately be considered as stable. At lower pion masses, the three-pion spectrum requires three pion operators which has only recently been done~\cite{Horz:2019rrn}. 

In this study, we compute the excited two and three-body spectrum of the multi-pion system at maximal isospin and compare it to the calculation by H\"orz and Hanlon~\cite{Horz:2019rrn}. The work is based on recent formal developments~\cite{Doring:2018xxx, Mai:2017bge}; we use the Inverse Amplitude Method (IAM) to one loop~\cite{Truong:1988zp, Dobado:1996ps, GomezNicola:2001as, GomezNicola:2007qj, Pelaez:2010fj, Nebreda:2010wv, Nebreda:2012ve} to predict the $I=2$ pion-pion $S$ and $D$-waves and then use the $S$-wave two-body input to predict the three-body finite-volume spectrum.  Several eigenvalues are calculated in moving frames~\cite{Horz:2019rrn} which requires us to extend our formalism to boosted frames.

While this paper was prepared for publication, an
independent study appeared~\cite{Blanton:2019vdk}
presenting a similar analysis of the $\pi^+$ spectra
generated by H\"orz and Hanlon~\cite{Horz:2019rrn}.

\section{Formalism}

\begin{figure}
    \centering
    \includegraphics[width=0.89\linewidth]{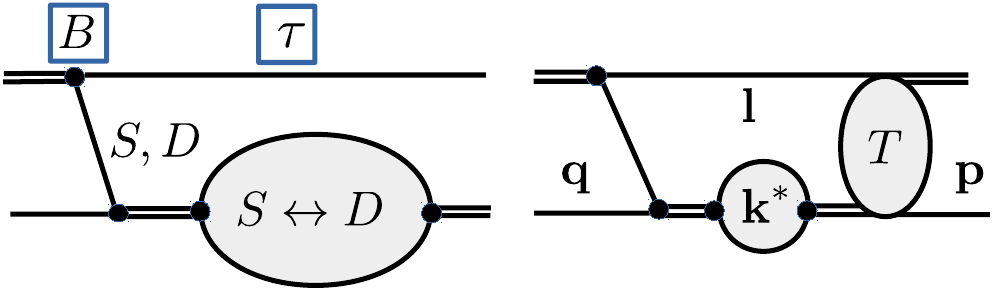}    
    \caption{Left:
    Partial waves in the isobar-spectator interaction and $S/D$ ``in flight'' transitions {forbidden in the infinite volume but allowed in finite volume} (schematically). Right: Momentum labeling of the three-body amplitude as used in the main text. The shown part corresponds to the second term in Eq.~(\ref{eq:finvolT}). }
    \label{fig:schemes}
\end{figure}

The three-body amplitude can be organized in the isobar-spectator picture; {to describe three-body on-shell states,} first, two particles are combined in terms of their quantum numbers and two-body interactions to form an isobar; the third particle, called spectator, is then added.
Using this parametrization, a relativistic three-body unitary amplitude was derived in Ref.~\cite{Mai:2017vot}.  This provides a complete proof of three-body unitarity above threshold missing in previous work~\cite{Aaron:1969my}. 
The amplitude is derived from dispersion relations, and can be matched to a Feynman diagrammatic approach but is a-priori independent of it. The isobar-spectator interaction itself is dictated by unitarity and develops an imaginary part. It can be represented as particle exchange as shown on the left-hand side in Fig.~\ref{fig:schemes}. 
There, solid lines indicate the spectator $\pi^+$ and double lines represent the isospin $I=2$ isobar; note that any two-body amplitude, as for example the repulsive $I=2,\ell=0$ can be mapped to the isobar picture~\cite{Mai:2018djl, Bedaque:1999vb}. 
In the present scheme, three-body forces arise naturally as real parts that can be added to the interaction without destroying unitarity. 

\begin{figure}
    \centering
    \includegraphics[width=0.99\linewidth]{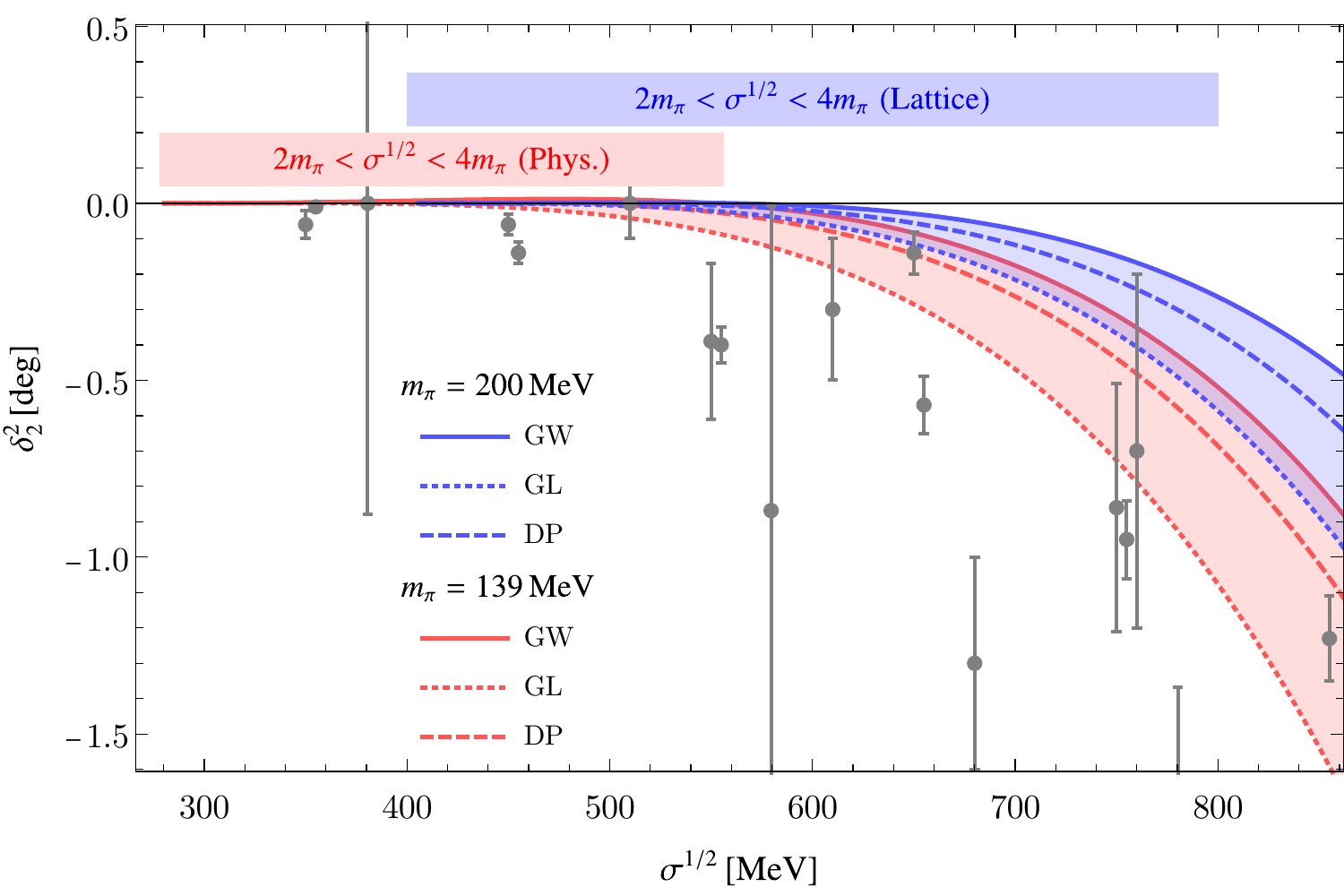}    
    \caption{Prediction of $D$-wave scattering at the physical point (red curves/area) compared to phase shifts extracted from experiment~\cite{Losty:1973et, Cohen:1973yx, Hoogland:1977kt}. For comparison, the predicted $D$-wave at the pion mass of Ref.~\cite{Horz:2019rrn} is also indicated (blue curves/area). The respective elastic regions are indicated with the horizontal bars. Predictions are shown using the LECs from Ref.~\cite{Mai:2019pqr} (GW), Ref.~\cite{Gasser:1983yg} (GL), and Ref.~\cite{Dobado:1996ps} (DP).    
    }
    \label{fig:Dwave}
\end{figure}

For the $3\pi^+$ system, the left-hand side of Fig.~\ref{fig:schemes} indicates that the $I=2$ $\pi\pi$ amplitude can only have even-spin isobars ($S, \, D, \dots$) due to Bose symmetry. Also, the isobar-spectator interaction can only be in even partial waves $\ell=0,\,2,\dots$. For both cases, we truncate the expansion at $D$-waves which is a good approximation for {low energies that is} backed by phenomenology~\cite{GarciaMartin:2011cn}. 
In addition, the $(I,\ell)=(2,2)$ $\pi\pi$ interaction is very small as shown in Fig.~\ref{fig:Dwave} for different low-energy constants (LECs).
Perturbative next-to-leading order (NLO) calculations (red curves and band) predict very small phase shifts not in contradiction with the scattered phase shifts from experiment. See also Ref.~\cite{Nebreda:2012ve} for a similar calculation, comparing also the LQCD phase shifts of Ref.~\cite{Dudek:2012gj}. If one chirally extrapolates the calculation to the pion mass of Ref.~\cite{Horz:2019rrn}, of $m_\pi\approx 200$~MeV (blue lines and band), one can see that the size of the $D$-wave stays below one degree in the elastic region.

We also find that there is no apparent sign of $D$-wave in the lattice data under consideration~\cite{Horz:2019rrn}.  In all irreducible representations (``irreps'') in which the $D$-wave is the lowest participating wave, the finite-volume energies coincide with noninteracting levels within uncertainties. For irreps with $S/D$-wave mixing, no {apparent} sign of $D$-wave is found, either, as discussed in the Results section. For our predictions, we will, therefore, neglect the $(I,\ell)=(2,2)$ $\pi\pi$ interaction in the following. However, there is no reason to exclude the relative $\pi^+$-isobar $D$-wave which will turn out to be important.

In Ref.~\cite{Mai:2017bge} the finite-volume version of the three-body amplitude  was derived, and, for the first time, the systematic cancellation of unphysical singularities and the practical applicability of a such a formalism was demonstrated, and correctly projected to the $A_1^-$ irrep. In Ref.~\cite{Mai:2018djl}, for the first time, a three-body formalism was compared to LQCD data of an actual physical system, $\pi^+\pi^+\pi^+$, including a fit of the three-body force. In summary, the only missing ingredient for the prediction of the new LQCD data consists in the development of a finite-volume formalism allowing for three-body systems in moving frames.

\subsection{Moving Three-body System}

For the formulation of the three-body $T$-matrix in finite volume~\cite{Mai:2017bge}, we took advantage of cubic symmetry which enabled us to arrange allowed lattice momenta on ``shells'' of equal absolute momenta. For moving three-body systems, cubic symmetry is broken and it is more advantageous to work in a three-dimensional momentum basis, suitably labeling the allowed momenta $\tilde {\bf r}_i=(2\pi/L)\, \tilde{\bf n}_i$ with $\tilde{\bf n}_i\in \mathds{Z}^3$. In the following, three-momenta with tilde are defined in the lattice rest frame, three-momenta without overscript are defined in the three-body rest frame, and starred three-momenta are defined in the two-body isobar rest frame. With this, the symmetrized three-body scattering amplitude in the three-body rest frame reads
\begin{align}
\label{eq:3bodyampl}
\langle q_1 q_2 q_3|\mathcal{T}|  p_1 p_2 p_3\rangle  
=
\frac{1}{3!}\sum_{n=1}^3\sum_{m=1}^3
v( q_{\bar{n}}, q_{\bar{\bar{n}}})
\hat T_{nm}(s)
v( p_{\bar{m}}, p_{\bar{\bar{m}}})\,,
\end{align}
where $(n,\bar n,\bar{\bar{n}})$ denotes a circular permutation of $(1,2,3)$ etc., and $v$ denotes the decay vertex of the isobar, which is chosen as specified in the appendix to reproduce exactly the Inverse Amplitude Method for the two-body sub-channel amplitudes~\cite{Mai:2018djl}. Note that this vertex also contains a smooth cutoff function which regulates all two and three-body integrals. This function is chosen as in Ref.~\cite{Mai:2018djl}, where it is shown that the dependence on the particular choice of the cutoff is very weak. The quantity $s$ represents the square of the total four momentum of the three-body system, such that the isobar-spectator amplitude $\hat T$ reads~\cite{Mai:2017bge} 
\begin{align}\label{eq:finvolThat}
\hat T_{nm}(s)&
=\tau_n(s)T_{nm}(s)\tau_m(s)
-2E_n L^3\tau_n(s)\delta_{nm} \ ,\\
\label{eq:finvolT}
T_{nm}(s)
&=
B_{nm}(s)
-\sum_{x}
\tilde {J}_x
B_{nx}(s)
\frac{\tau_{x}(s)}{2L^3E_x}
T_{xm}(s)\,,
\end{align}
where $m,n,x$ label the incoming spectator momentum ${\bf p}_m$, outgoing spectator momentum ${\bf q}_n$, and intermediate spectator momentum ${\bf l}_x$. A graphical representation of the second (``rescattering'') term of Eq.~(\ref{eq:finvolT}) is given on the right-hand side of Fig.~\ref{fig:schemes}. Furthermore,  $E_n=\sqrt{m_\pi^2+{\bf q}_n^2}$ and analogously for the other momenta. The Jacobian for the mapping from the lattice frame to the three-body rest frame is denoted by $\tilde {J}_x$. The quantities $B$ (``exchange+three-body'') and $\tau$ (``three-body propagation'') are defined in the Appendix and graphically indicated in Fig.~\ref{fig:schemes} to the left (up to a real three-body term, see Appendix). The projection to irreps is detailed in the Appendix, as well. 

In summary, for incoming and outgoing spectator momenta $\tilde {\bf p}_i$ and $\tilde {\bf q}_i$, the $3\pi^+$ system has  momentum 
$\tilde{\bf P}=\tilde{\bf q}_1+\tilde{\bf q}_2+\tilde{\bf q}_3=
\tilde{\bf p}_1+\tilde{\bf p}_2+\tilde{\bf p}_3$. A boost of lattice momenta by $\tilde{\bf P}$ provides the three-momenta entering Eqs.~(\ref{eq:finvolThat}, \ref{eq:finvolT}) that is solved in the three-body rest frame; another boost to the isobar rest frame is necessary as the pertinent summations are carried out in that frame. Schematically, we can represent this two-step process as follows:
\begin{center}
\includegraphics[width=0.99\linewidth]{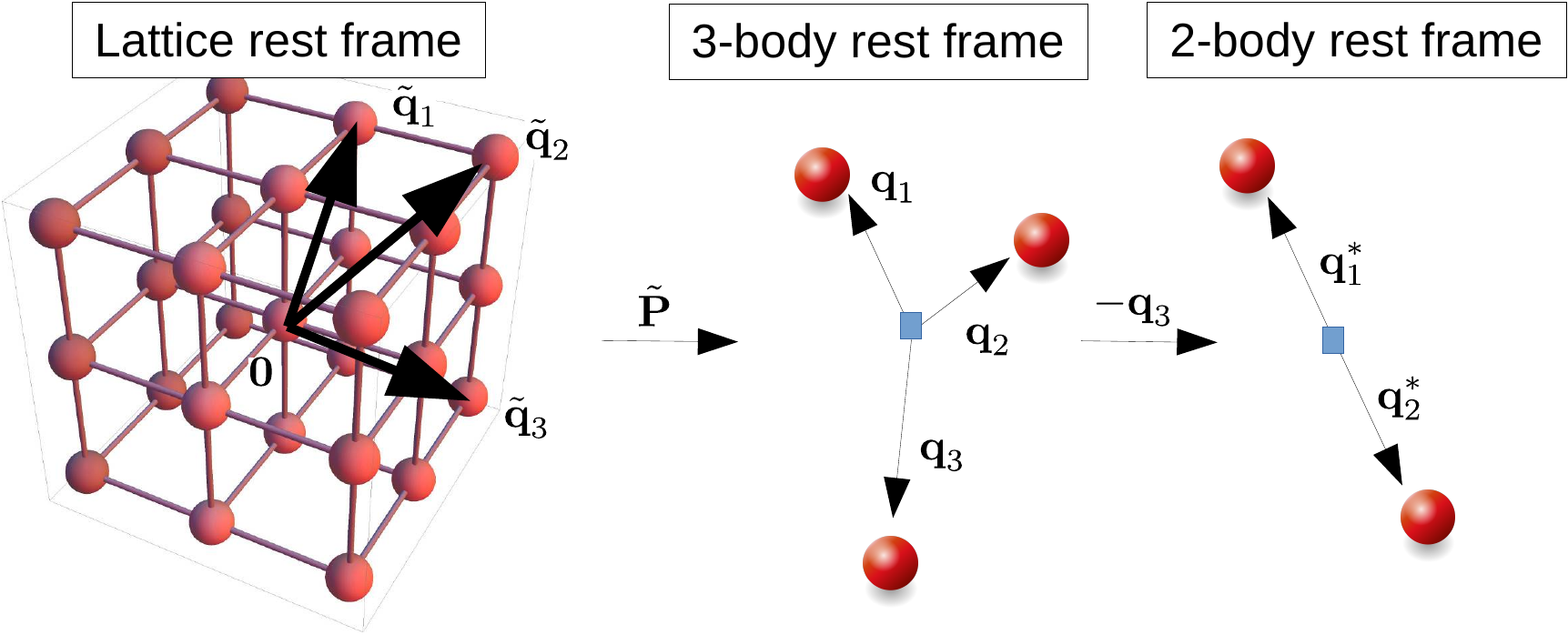}  
\end{center}
As a result of the formalism (see Appendix), the three-body system in moving frames is entirely expressed in terms of lattice momenta $\tilde{\bf p}_m,\,\tilde{\bf q}_n\in (2\pi/L)\mathds{Z}^3$, and the invariant $s$. Its poles indicate the energy eigenvalues after projection to irreps.

\section{Results}

\begin{figure*}
    \centering
    \includegraphics[width=0.99\linewidth]{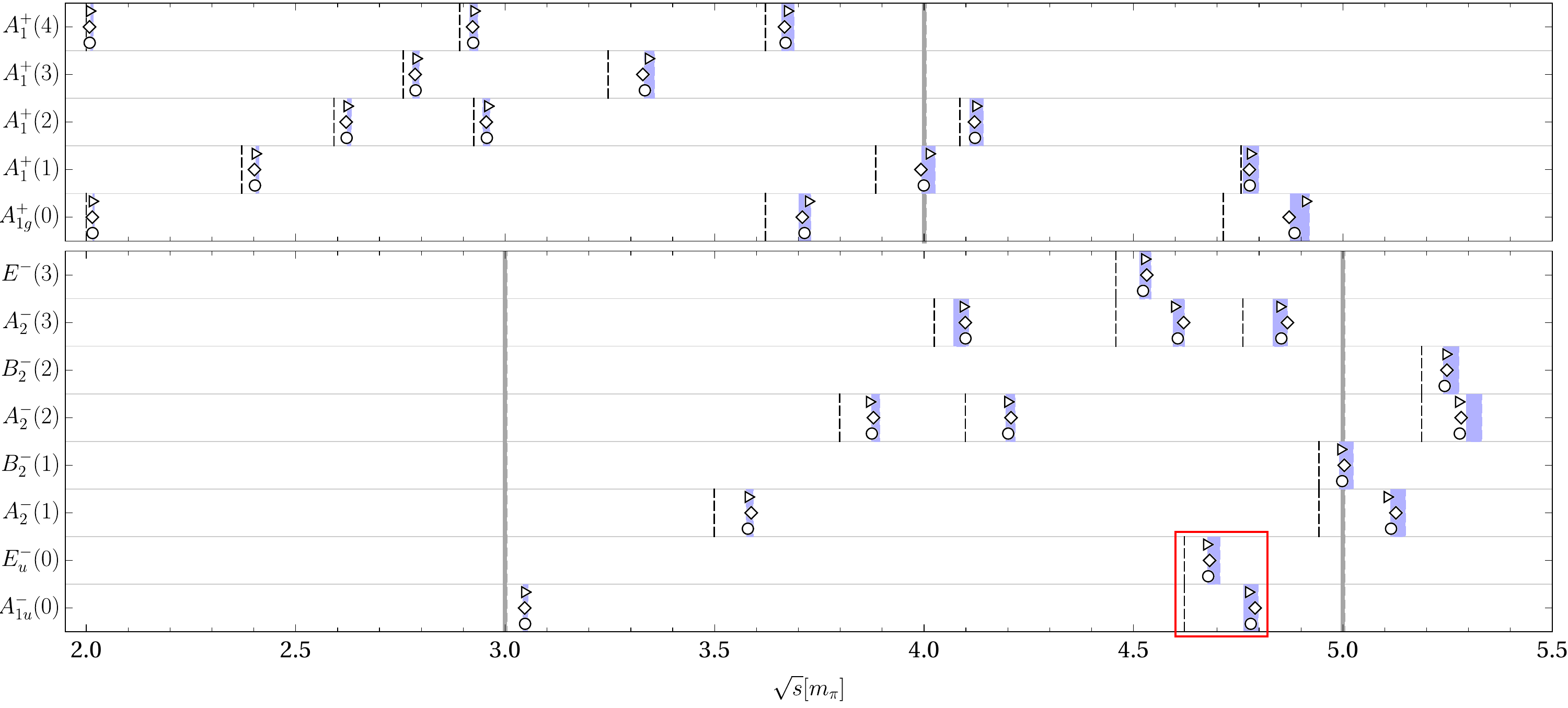}    
    \caption{Top: Prediction of the two-body spectrum for irreps in which the $S$-wave participates. The $\pi\pi$ $D$-wave is set to zero as discussed in the text. The lattice data~\cite{Horz:2019rrn} are represented with the small shaded bars and chiral predictions with the symbols, depending on different values for low-energy constants from Ref.~\cite{Nebreda:2012ve} ($\triangleright$), Ref.~\cite{Mai:2019pqr} ($\Diamond$, based on lattice results of the GW group~\cite{Guo:2016zos, Guo:2018zss, Culver:2019qtx}), and Ref.~\cite{Gasser:1983yg} ($\ocircle$). Bottom: Predictions for the three-body sector for the same choice of LECs and vanishing three-body force. Note that the upper indices of the irreps indicate the $G$-parity, and the values in parentheses show the size of the respective boost,  following the notation of Ref.~\cite{Horz:2019rrn}.}
    \label{fig:hoerzpred2b}
\end{figure*}

\begin{figure}
    \centering
    \includegraphics[width=0.85\linewidth]{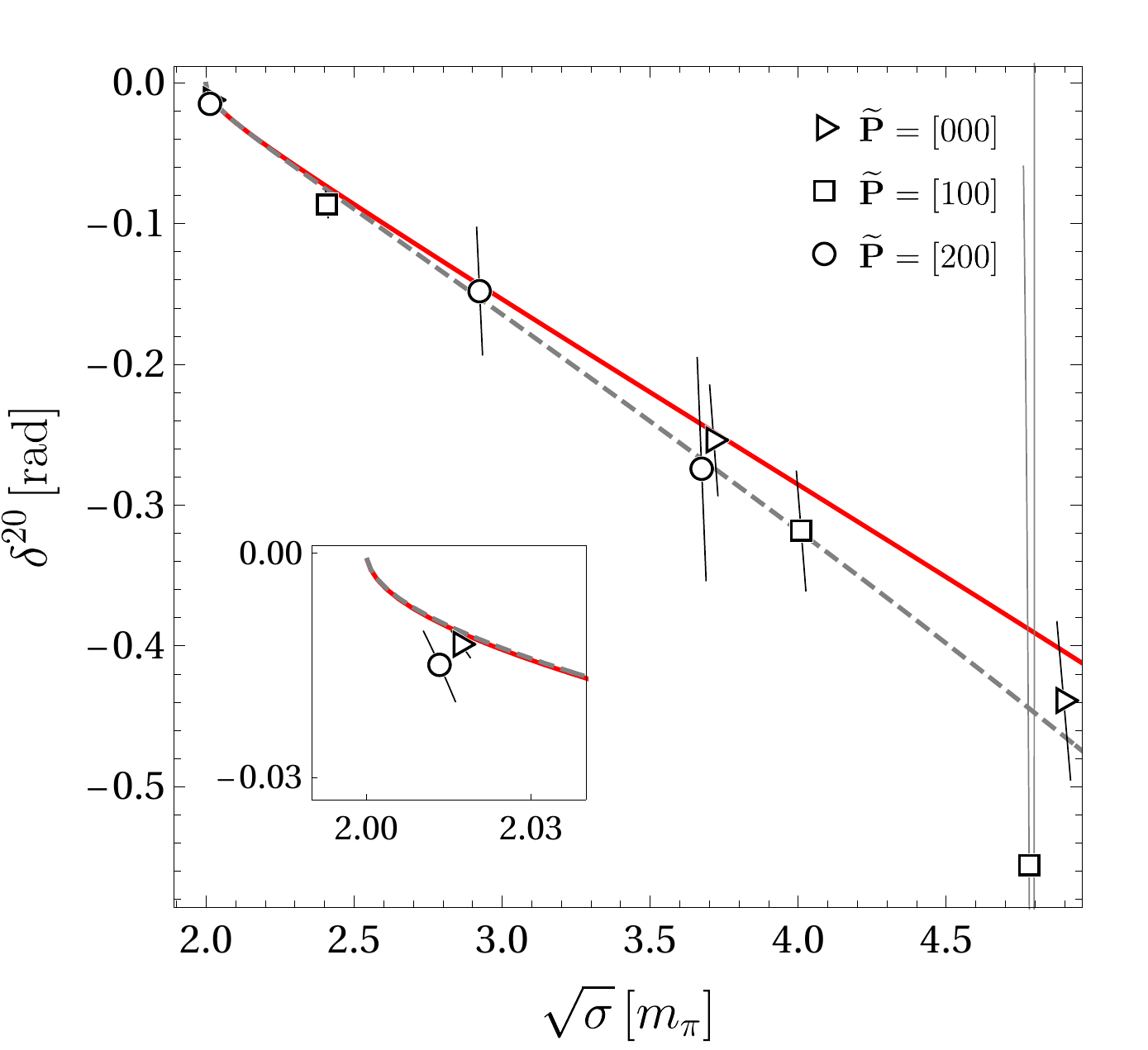}    
    \caption{Predictions of the $I=2$ $S$-wave phase shift for low-energy constants and their uncertainties from Ref.~\cite{Gasser:1983yg} (red solid line) and Ref.~\cite{Nebreda:2012ve} (gray dashed line). For comparison, we also extract some phases from the eigenvalues of Ref.~\cite{Horz:2019rrn} (data points).}
    \label{fig:hoerzphases}
\end{figure}

Taking the two-body input from IAM~\cite{Truong:1988zp, Dobado:1996ps, GomezNicola:2001as, GomezNicola:2007qj, Pelaez:2010fj, Nebreda:2010wv, Nebreda:2012ve}
we predict the energy eigenvalues for the $\pi^+\pi^+$ and $\pi^+\pi^+\pi^+$  systems.
The uncertainties for our prediction are estimated by using central values for LECs from three different analyses~\cite{Nebreda:2012ve, Mai:2019pqr, Gasser:1983yg} instead of LEC uncertainties. The reason is that, usually, no correlations on LECs are quoted in the literature which leads to an uncontrolled overestimation of the prediction error.
For the LECs of Ref.~\cite{Gasser:1983yg}, results are quoted in Tab.~\ref{tab:energeigenvalues2and3}. For all results, the $D$-wave isobar is neglected as discussed before, and the three-body term is set to zero, $C=0$ (see discussion below). 
The predictions for the two-body (three-body) spectrum are represented in the upper (lower) part of Fig.~\ref{fig:hoerzpred2b}.
For some of the irreps, phases are extracted and shown together with chiral predictions in Fig.~\ref{fig:hoerzphases} for illustration.
Overall, the predictions from different LECs vary surprisingly little given the different origins of their determination. Furthermore, the predictions are all quantitatively very good.  

In all $A_1^+$ irreps with non-vanishing boost, $S$ and $D$-waves mix in the $\pi^+\pi^+$ system. At higher energies, one could therefore expect deviations of the predictions from the data as we have neglected the $D$-wave throughout. However, the quality of our predictions, even beyond the elastic threshold,
adds another piece of evidence that the $D$-wave can be neglected. Of course, only a $S/D$-coupled partial-wave fit can provide ultimate clarity for this point (see, e.g., Refs.~\cite{Dudek:2012gj, Doring:2012eu}). In Ref.~\cite{Blanton:2019vdk}, some evidence for a non-vanishing $D$-wave was found by fitting only irreps in which the lowest participating wave is the $D$-wave.

The quality of our predictions can be assessed by evaluating the correlated $\chi^2/n$ with $n$ being the number of lattice eigenvalues in the respective elastic regions. For LECs set to the values of Ref.~\cite{Gasser:1983yg} we have
\begin{align}
\frac{\chi_{(2)}^2}{n}\approx\frac{21.2}{11}\ , \quad
\frac{\chi_{(3)}^2}{n}\approx
\frac{9.5}{11} \ ,
\quad
\frac{\chi_{(2\&3)}^2}{n}\approx
\frac{39.4}{22} \,
\nonumber
\end{align}
for the two-, three-, and combined two and three-pion sectors, including the cross-correlations of energy eigenvalues. The $\chi^2$ values for the LECs from Refs.~\cite{Nebreda:2012ve, Mai:2019pqr} are very similar to the quoted ones.

The predictions were obtained with a vanishing three-body force $C$. In Ref.~\cite{Mai:2018djl}, $C$ was fitted to the ground-state level and found to be negligible. However, the present data~\cite{Horz:2019rrn} are more precise than the NPLQCD data~\cite{Detmold:2008fn}.  Any deviation of the prediction in the three-body sector, especially at higher energies, could be a sign for a non-vanishing three-body force at the chosen regularization.
This is obviously not the case {as $\chi^2_{(3)}/n\approx 1$}. Moreover, a large part of the overall $\chi_{(2\&3)}^2\approx 39.4$ arises from the correlations of one point at low energies  ($\sigma^{1/2}\approx 2.4\,m_\pi$,  $\pi^+\pi^+$ sector, $A_1^+(1)$) with the $3\pi^+$ sector. Without this point, $\chi_{(2\&3)}^2/n\approx28/21\approx 1.3$ and it is difficult to explain this change with the discussed simplifications of our formulation.

To conclude, consider the excited-state $3\pi^+$ energy shifts in $A_{1u}^-(0)$ and $E_u^-(0)$ highlighted in Fig.~\ref{fig:hoerzpred2b}. The relative and absolute sizes of these shifts are governed by
the structure of the exchange term $B$ shown in Fig.~\ref{fig:schemes} because that term determines the strength of S-wave vs. D-wave interactions. On the other hand, that term
arises as a consequence of three-body unitarity~\cite{Mai:2017wdv} and contributes to the powerlaw finite-volume effects~\cite{Mai:2017bge}. 
In conclusion, for the first time, three-body unitarity is directly visible in LQCD data. This conclusion would hold similarly  for any parametrization of the two-body sector, i.e., it is independent of the IAM model we choose for our predictions. 

\section{Conclusions}

Using a two-body unitary amplitude, that matches Chiral Perturbation Theory up to next-to-leading order (IAM), the isospin $I=2$ two-body body eigenvalues of a recent lattice QCD calculation~\cite{Blanton:2019vdk} were predicted in a restriction to $S$-wave. With this two-body input, three-body unitarity served as the $S$-matrix constraint to predict the three-body spectrum with a correlated $\chi^2_{(3)}/n\approx 10/11$, i.e., no sign of a substantial three-body force was seen for the given regularization. Yet, if correlations of the two- and three-body sector are combined, a $\chi^2_{(2\&3)}/n\approx 1.8$ indicates a residual tension. We want to stress that the LECs are not fit to the lattice data; the tension is likely to disappear if we adjust the LECs to minimize $\chi^2$.
Overall, the predictions, depending only on low-energy constants from independent studies (and, very weakly, on the regularization), are in good agreement with the data. Furthermore, the correct prediction of the $S$-wave and $D$-wave excited-level energy shifts in $A_{1u}^-(0)$ and $E_u^-(0)$ depends only on the structure of the spectator-isobar interaction, which, in turn, is dictated by three-body unitarity. For the first time, this fundamental $S$-matrix principle is directly visible in lattice QCD data.

\bigskip

\begin{acknowledgments}
MD and MM acknowledge support by the National Science Foundation (grant no. PHY-1452055) and by the U.S. Department of Energy, Office of Science, Office of Nuclear Physics under grant no. DE-SC0016582. AA and CC are supported in part by U.S. Department of Energy, Office of Science, Office of Nuclear Physics under grant no DE-FG02-95ER40907.
\end{acknowledgments}

\bibliography{ALL-REF.bib}
\begin{onecolumngrid}
\appendix
\section{Formalism for Moving Frames}

The necessary formalism to calculate eigenvalues for three-body systems in moving frames is provided. As discussed in the main text, for incoming and outgoing momenta $\tilde {\bf q}_i$ and $\tilde {\bf p}_i$, the $3\pi^+$ system has  momentum $\tilde{\bf P}=\tilde{\bf q}_1+\tilde{\bf q}_2+\tilde{\bf q}_3=
\tilde{\bf p}_1+\tilde{\bf p}_2+\tilde{\bf p}_3$, 
where $\tilde{\bf P}\in (2\pi/L) \{(0,0,1),\,(0,1,1),\,(1,1,1)\}$ and multiples thereof. The momenta in the three-body rest frame are~\cite{Doring:2012eu}
\begin{align}
{\bf q} = \tilde{\bf q}+\left[
\left(\frac{\tilde P^{0}}{\sqrt{s}}-1\right)\frac{\tilde{\bf q}\tilde{\bf P}}{|\tilde{\bf P}^2|}-\frac{\tilde q^{\,0}}{\sqrt{s}}\right]{\bf \tilde P}\,,
\label{moveframe}
\end{align}
and analogously for the other momenta ${\bf p}$ and ${\bf l}$.
In Eq.~(\ref{moveframe}), $\tilde q^{\,0}=\sqrt{\tilde {\bf q}^2+m_\pi^2}$ and $\tilde P^0=\sqrt{s+\tilde{\bf P}^2}$, see Ref.~\cite{Doring:2012eu}.  
For a finite boost, the Jacobian appearing in Eq.~(\ref{eq:finvolT}) of the main text is evaluated from Eq.~(\ref{moveframe}) as
\begin{align}
\tilde J_x=
\left|\frac{dl_i}{d\tilde l_j}\right|
=
\frac{\tilde P^0}{\sqrt{s}}-\frac{\tilde {\bf l}_x\tilde{\bf P}}{\sqrt{s}\,\tilde l^0_x}
\label{Jacobian}
\end{align}
with $\tilde l^0_x=\sqrt{m_\pi^2+\tilde {\bf l}_x^2}$. The isobar is not at rest in the three-body rest frame. Thus, an additional boost (by $-{\bf l}$) has to be performed for the pertinent summation of momenta ${\bf k}^*$ in the self energy of the isobar. This is detailed in Eqs.~(11, 12) of Ref.~\cite{Mai:2017bge} and reads in the current notation
\begin{align}\label{eq:boost}
\boldsymbol{k}^*(\boldsymbol{k},\boldsymbol{l}_m)
=
\boldsymbol{k}+
\boldsymbol{l}_m\Bigg(\frac{\boldsymbol{k}\cdot \boldsymbol{l}_m}{\boldsymbol{l}_m^2}\Big(\frac{\sqrt{\sigma_m}}{\sqrt{s}-l_m^0}-1\Big)+
\frac{\sqrt{\sigma_m}}{2(\sqrt{s}-l_m^0)}\Bigg)
\text{~~with~~}
J_m=\frac{\sqrt{\sigma_m}}{\sqrt{s}-l_m^0}\,,
\end{align}
denoting the corresponding Jacobian. The quantity $\sigma_m=s+m_\pi^2-2\sqrt{s}l_m^0$ is the square of the invariant mass and $l_m^0=\sqrt{m_\pi^2+{\bf l}_m^2}$. The isobar propagator in Eqs.~(\ref{eq:finvolThat}, \ref{eq:finvolT}) of the main text reads then
\begin{align}\label{eq:finvolTAU}
\tau_{m}^{-1}(s)
&=
\sigma_m-M_0^2
-\frac{1}{L^3}
\sum_i
\frac{\tilde J_m\,J_m\,
\left(\lambda(\sigma_m) f(4(\boldsymbol{ k}^*_i)^2)\right)^2}
{2k_i^{0*}
\left(\sigma_m-4\left(k_i^{0*}\right)^2\right)} \ ,
\text{~~where~~}
\boldsymbol{k}_i^*\equiv
\boldsymbol{k}^*(\boldsymbol{k}(\tilde{\boldsymbol{k}}_i),\boldsymbol{l}_m) \ ,
\end{align}
where
$\tilde{\boldsymbol{k}}_i\in (2\pi/L) \mathds{Z}^3$ and $k_i^{0*}=\sqrt{m_\pi^2+\boldsymbol{k}_i^{*2}}$.
The numerator in Eq.~(\ref{eq:finvolTAU}) contains also the isobar  $S$-wave decay vertex $v=\lambda f$ with a form-factor $f$, which regulates the appearing integrations/summations over momenta. Following the discussion of Ref.~\cite{Mai:2018djl}, we choose $f(Q^2)=1/(1+e^{-(\Lambda/2-1)^2+(Q/m_\pi)^2/4})$ with $\Lambda=42$. The dependence of the results on 
$\Lambda$ has been checked thoroughly in Ref.~\cite{Mai:2018djl} and 
was found to be very mild. Similarly, various analytic forms of the form factor have been evaluated and compared in the same publication. 
Furthermore, 
the matching to NLO IAM is expressed as~\cite{Mai:2018djl}
\begin{align}
\lambda(\sigma)^2=(M_0^2-\sigma)\Big(\frac{d}{4\pi^2}+\frac{T_{LO}-\bar T_{NLO}}{T^2_{LO}}\Big)^{-1}\,,
\end{align}
where $T_{LO}$ is the leading-order chiral $\pi\pi$ scattering amplitude, and $\bar T_{NLO}$ denotes the next-to-leading order amplitude without the $s$-channel loop. The latter part depends on four LECs, which are fixed as discussed in the main body of the paper. The parameter $d=0.86$ makes a connection between the regularization by form factors (performed in this work) and the dimensional regularization on the level of $\pi\pi$ scattering amplitudes. This matching is necessary due to the fact that we use the LECs extracted in the latter scheme. Further details on this technicality are discussed in the Ref.~\cite{Mai:2018djl}. Overall, the above choice of the coupling $\lambda$ leads to the form of the two-body sub-channel amplitudes, which match the Inverse Amplitude Method~\cite{Truong:1988zp,Pelaez:2015qba}. This type of amplitudes matches the ChPT amplitude up to the next-to-leading order exactly, allowing also for addressing all three isospin channels of the $\pi\pi$ system in a large energy region as recently demonstrated in Ref.~\cite{Mai:2019pqr}.

For completeness, we also quote the (unprojected) driving term
\begin{align}\label{eq:bfinvol}
B_{nm}(s)
=
-&
\frac{\lambda^2f((\sqrt{s}-2E_m-E_n)^2-|2\boldsymbol{p}_{m}+\boldsymbol{q}_{n}|^2)
f((\sqrt{s}-2E_n-E_m)^2-|2\boldsymbol q_{n}+\boldsymbol{p}_{m}|^2)}
{2E_{\text{ex}}
(\sqrt{s}-E_m-E_n-E_{\text{ex}})}
-C(\boldsymbol{q}_{n},\boldsymbol{p}_{m};s) 
\ ,
\end{align}
where $E_{\text{ex}}^2=m_\pi^2+(\boldsymbol{q}_{n}+\boldsymbol{p}_{m})^2$. With all parts of Eqs.~(\ref{eq:finvolThat}) of the main text defined, the $\hat T$ matrix can be calculated; its poles coincide with the three-body energy eigenvalues in moving frames.

\subsection{Predicted Energy Eigenvalues}

In the two-body sector, the positions of the poles of the two-body scattering amplitude $T_{22}=v \tau v$ give the two-body energy eigenvalues in the $A_1^+$ irrep. In the three-body case, the projections to the corresponding irreps is performed similarly to the method of Refs.~\cite{Horz:2019rrn, Morningstar:2013bda}, see Ref.~\cite{Doring:2018xxx}. In particular for Eq.~\eqref{eq:3bodyampl} of the main text,
\begin{align}
    \mathcal{T}^{\Gamma}(s)=\sum_{i,j}\chi^\Gamma(R_i)\chi^\Gamma(R_j)  \langle R_j q_{1,2,3}|\mathcal{T}(s)|R_j p_{1,2,3}\rangle\,,
\end{align}
where the indices $i$ and $j$ run over all group elements and the coefficients $\chi$ are the characters of the group elements, see, e.g., Refs.~\cite{Horz:2019rrn, Morningstar:2013bda}. Here, $\Gamma$ denotes the irreps $A_{1u}^-$, $E_{u}^-$, $A_{2}^-$, $B_{2}^-$, and $E^-$.  The predicted energy eigenvalues are shown in Table~\ref{tab:energeigenvalues2and3}.

\renewcommand{\arraystretch}{1.1}
\begin{table}[ht]
    \begin{ruledtabular}
    \begin{tabular}{c|l|lllll}
        & \multicolumn{1}{c|}{two-body}    
        & \multicolumn{5}{c}{three-body}\\
    \hline  
        \addlinespace[0.15em]
    $\tilde{\bf P}$   
    &$A_1^+$
    &$A_{1u}^-$ & $E_u^-$ & $A_2^-$ & $B_2^-$ & $E^-$
    \\
            \addlinespace[0.1em]
    \hline
            \addlinespace[0.1em]
    $[000]$     
    &2.015~3.715~4.885
    &3.048~4.780    &4.679  & &&\\
    $[100]$     
    &2.403~3.999~4.778
    & & &3.579~5.115  &4.998 &  \\
    $[110]$     
    &2.622~2.957~4.122
    &&&3.876~4.201~5.279&5.243 &\\
    $[111]$     
    &2.786~3.334 
    &&&4.099~4.606~4.853 & &4.523    \\
    $[200]$     
    &2.008~2.924~3.670 
    &&&&& 
    \end{tabular}
    \end{ruledtabular}
    \caption{\label{tab:energeigenvalues2and3}
    Predictions of two and three-body finite-volume eigenvalues using LECs from Ref.~\cite{Gasser:1983yg}. For notation, see caption of Fig.~\ref{fig:hoerzpred2b}.}
\end{table}
\end{onecolumngrid}
\end{document}